\documentclass[prl,aps,showpacs,superscriptaddress,twocolumn]{revtex4}
\usepackage{graphicx}
\usepackage{amsmath}
\usepackage{amsfonts}
\usepackage{amssymb}
\usepackage{epsfig}

\begin{document}
\title{
Scaling of Berry's Phase Close to the Dicke Quantum Phase
Transition}
\author{Francesco Plastina}\affiliation{Dipartimento di Fisica, Universit\`{a} della Calabria, Italy and
Istituto Nazionale di Fisica Nucleare, Gruppo collegato di
Cosenza, I-87036 Arcavacata di Rende, Cosenza, Italy}
\author{Giuseppe Liberti}\affiliation{Dipartimento di Fisica, Universit\`{a} della Calabria, Italy and
Istituto Nazionale di Fisica Nucleare, Gruppo collegato di
Cosenza, I-87036 Arcavacata di Rende, Cosenza, Italy}
\author{Angelo Carollo}\affiliation{
Institute for Quantum Optics and Quantum Information of the
Austrian Academy of Sciences, A-6020 Innsbruck, Austria}

\begin{abstract}
We discuss the thermodynamic and finite size scaling properties of
the geometric phase in the adiabatic Dicke model, describing the
super-radiant phase transition for an $N$ qubit register coupled
to a slow oscillator mode. We show that, in the thermodynamic
limit, a non zero Berry phase is obtained only if a path in
parameter space is followed that encircles the critical point.
Furthermore, we investigate the precursors of this critical
behavior for a system with finite size and obtain the leading
order in the $1/N$ expansion of the Berry phase and its critical
exponent.
\end{abstract}

\date{\today}

 \maketitle A considerable understanding
of the formal description of quantum mechanics has been achieved
after Berry's discovery~\cite{Berry84} of a geometric feature
related to the motion of a quantum system. He showed that the wave
function of a quantum system retains a memory of its evolution in
its complex phase, which only depends on the ``geometry'' of the
path traversed by the system. Known as the \emph{geometric phase},
this contribution originates from the very heart of the structure
of quantum mechanics. Historically, the first implicit derivation
of geometric phase is to be found in the work of H. C.
Longuet-Higgins et al.~\cite{Longuet-HOPS58} and G. Herzberg and
H. C. Longuet-Higgins~\cite{HerzbergL-H63} on the study of the
$E\otimes e$ Jahn-Teller problem in molecular systems. They showed
that the Born-Oppenheimer wave function, if required to be real
and smoothly varying, undergoes a sign change (a $\pi$ geometric
phase shift) when the nuclear configuration is brought around a
crossing point of the energy levels, i.e. a point where the energy
levels become degenerate. The widespread interest in geometric
phases is unquestionably due to the work of Berry~\cite{Berry84}
who put the concept of geometric phase in an unexpectedly general
scenario, showing that it is a property of the evolution of any
quantum mechanical system. It is then not too surprising how this
idea has been so broadly studied and applied to a variety of
contexts~\cite{ShapereW89,BohmMKNZ03}.

An apparently unrelated area in which a connection with Berry
Phase (BP) has been recently drawn is the study of Quantum Phase
Transitions (QPT) in many-body
systems~\cite{CarolloP05L,Zhu06L,Hamma05X}. It is a well known
fact that QPT~\cite{Sachdev99} are accompanied by a qualitative
change in the nature of correlations in the ground state of a
quantum system, and describing these changes is clearly one of the
major interests in condensed matter physics. In particular,
critical points are associated with a non-analytical behavior  of
the system energy density~\cite{Sachdev99} . It is expected that
such drastic changes are reflected in the geometry of the Hilbert
space. The geometric phase is able to capture singular behaviors
of the wave function, and is therefore expected to signal the
presence of QPT. The appearance of non-trivial BP in presence of
criticality is also heuristically motivated by the fact that QPT
arise in correspondence of energy level crossings (in the
thermodynamical limit) between ground and excited
states~\cite{Sachdev99}. As implicitly suggested by the early work
of H. C. Longuet-Higgins et al.~\cite{Longuet-HOPS58}, level
crossings generate singularities in the curvature of the Hilbert
space.  Having the ability to adiabatically change the parameters
of the system so as to encircle these singularities, can reveal
the critical dependence of geometric
phases~\cite{CarolloP05L,Hamma05X}. Indeed, due to its topological
character, the BP provides a tool to detect and probe
criticalities without taking the system through a phase
transition. This has been recently considered in
ref.~\cite{CarolloP05L}, where a (1+1) XY spin model is shown to
display a topological behavior of the BP when a path that
circulates (or not) a region of XX criticality is taken. Zhu have analyzed
further the connection between BP and QPT, obtaining a scaling
relation for the geometric phase~\cite{Zhu06L}. Very recently,
Hamma has demonstrated the topological character of the BP across
regions of QPT for a general model of many body
system~\cite{Hamma05X}.

These results have been achieved in the thermodynamic limit and it
would be interesting, in this context, to understand whether and
how precursors of QPT for finite size systems appear in the
geometric phase. In this letter, we analyze this problem for the
case of the critical behavior of the Dicke Model (DM) and
investigate the topological and scaling properties of the
geometric phase across a region of criticality. We consider a
system which consists of $N$ two-level systems (a qubit register
or an ensamble of indistinguishable atoms) coupled to a single
oscillator (bosonic) mode. The Hamiltonian is given by (in unit
such that $\hbar=c=1$)
\begin{equation}
    H={\omega} a^{\dagger}a+\Delta S_x+
   \frac{\lambda}{\sqrt{N}}(a^{\dagger}+a)S_z
    \label{1}
    \end{equation}
where $\Delta$ is the transition frequency of the qubit, $\omega$
is the frequency of the oscillator and $\lambda$ is the coupling
strength. The qubit operators are expressed in terms of total spin
components ${S}_{k}=\sum_{i=1}^{N}{\sigma}_{k}^{i}$, where the
${\sigma}_{k}^{i}$ 's ($k=x,y,z$) are the Pauli matrices used to
describe the $i$-th qubit.
%These operators obey the usual
%commutation relations $[S_k,S_l]=2i\epsilon_{klm} S_m$.
A $\pi/2$ rotation around the $y$ axis shows that $H$ is
canonically equivalent to the Dicke Hamiltonian \cite{dicke},
including counter-rotating terms.

After the first derivation due to Hepp and Lieb \cite{hepp}, the
thermodynamic properties of the DM have been studied by many
authors\cite{WH,Duncan,gilmore,orszag,widom,liberti}. In the
thermodynamic limit ($N\rightarrow \infty$), the system exhibits a
second-order phase transition at the critical point
$\lambda_c=\sqrt{\Delta\omega/2}$, where the ground state changes
from a normal to a super-radiant phase in which both the field
occupation and the spin magnetization acquire macroscopic values.
The continued interest in DM stems from the fact that it displays
a rich dynamics where many non-classical effects have been
predicted \cite{milburn,brandes,frasca,Hou,orszagent}, and from
its broad range of applications \cite{phyrep}. Investigations of
the ground state entanglement of the DM have been recently
performed \cite{lambert,reslen,vidal}, pointing out a scaling
behavior around the critical point.

The aim of present work is to investigate the DM in the adiabatic
regime ($\Delta \gg \omega$) and to demonstrate the topological
character of the BP for this case. Furthermore, we show that the
geometric phase obeys scaling relation at the critical point for a
system with finite size

In order to generate a Berry phase we change the Hamiltonian by
means of the unitary transformation:
\begin{equation}\label{ut}
    U(\phi)=\exp{\left(-i\frac{\phi}{2}S_x\right)}
\end{equation}
where $\phi$ is a slowly varying parameter, changing from $0$ to
$2 \pi$.

The transformed Hamiltonian can be written as
\begin{equation}
   {H}(\phi)=U^\dag(\phi) H U(\phi)=\frac{\omega}{2}\left[p^2+q^2+\mathbf{G}\cdot
   \mathbf{S}\right]
    \label{ht2}
\end{equation}
where $\mathbf{G}=\left(D,\frac{L q}{\sqrt{N}}\sin{\phi},\frac{L
q}{\sqrt{N}}\cos{\phi}\right)$ is an effective vector field. Here,
 $D=2\Delta/\omega$ and $L=2\sqrt{2}
\lambda / \omega$ are dimensionless parameters and the Hamiltonian
of the free oscillator field is expressed in terms of canonical
variables $q =(a^{\dagger}+a)/\sqrt{2}$ and
$p=i(a^{\dagger}-a)/\sqrt{2}$ that obey the quantization condition
$[q,p]=i$.

In the adiabatic limit \cite{adiabatic,finite}, where we assume a
{\it slow} oscillator and work in the regime $D\gg1$, the
Born-Oppenheimer approximation can be employed to write the ground
state of $H(\phi)$ as:
\begin{equation}\label{deco}
    |\psi_{tot}\rangle=\int d q \, \varphi(q) |q\rangle \otimes |\chi(q,\phi)\rangle
\end{equation}
Here, $|\chi (q,\phi)\rangle$ is the state of the ``fast
component''; namely, the lowest eigenstate of the ``adiabatic''
equation for the qubit part, displaying a parametric dependence on
the slow oscillator variable $q$,
\begin{equation}\label{adiaham}
    \mathbf{G}\cdot \mathbf{S}|\chi(q,\phi)\rangle=E_l(q)|\chi (q,\phi)\rangle \,,
\end{equation}
As the qubits are indistinguishable, it is easy to prove that the
ground state can be expressed as a direct product of $N$ identical
factors,
\begin{equation}\label{qubitstates}
|\chi(q,\phi)\rangle=|\chi (q,\phi)\rangle_1\otimes |\chi
(q,\phi)\rangle_2\otimes\dots\otimes|\chi (q,\phi)\rangle_N
\end{equation}
Each component can be written as
\begin{equation}\label{gsdgen}
|\chi (q,\phi)\rangle_i=
\sin{\frac{\beta}{2}}|\uparrow\rangle_i-\cos{\frac{\beta}{2}}e^{i\zeta}
|\downarrow\rangle_i, \end{equation} where $|\uparrow\rangle_i$
and $|\downarrow\rangle_i$ are the eigenstates of $\sigma_i^z$
with eigenvalues $\pm1$, and where
\begin{eqnarray}\label{gsdgen2}
&& \cos{\beta} = {{\frac{Lq\cos{\phi}}{\sqrt{N} E(q)}}}\\
&& \zeta = \arctan{\frac{L q \sin{\phi}}{\sqrt{N} D}}
\end{eqnarray}
Here, $E(q)$ is related to the energy eigenvalue of Eq.
(\ref{adiaham}) as
\begin{equation}\label{eival}
    E_{l}(q)= - N E(q)=- N \sqrt{D^2+\frac{L^2q^2}{N}}\,.
\end{equation}
In the Born-Oppenheimer approach, this energy eigenvalue
constitutes an effective adiabatic potential felt by the
oscillator together with the original square term:
\begin{equation} V_{l}(q)=\frac{\omega}{2}\left[q^2- N E(q)\right].\end{equation}
Introducing the dimensionless parameter $ \alpha={L^2}/{2 D}$, one
can show that for $\alpha\leq1$, the potential $V_{l}(q)$ can be
viewed as a broadened harmonic well with minimum at $q=0$ and
$V_{l}(0)=-N\Delta$. For $\alpha>1$, on the other hand, the
coupling with the qubit splits the oscillator potential producing
a symmetric double well with minima at $\pm
q_m=\pm\frac{\sqrt{N}D}{L}\sqrt{\alpha^2-1}$ with $
V_{l}(q_m)=-\frac{N\Delta}{2}\left(\alpha+\frac{1}{\alpha}\right)$.

As the last step in the Born-Oppenheimer procedure, we need to
evaluate the ground state wave function for the oscillator,
$\phi_0(q)$, that has to be inserted in Eq. (\ref{deco}) to obtain
the ground state of the composite system. This wave function is
the normalized solution of the one-dimensional time independent
Schr\"odinger equation
\begin{equation}\label{se}
H_{ad}\varphi_{0}(q)=\left(-\frac{\omega}{2}\frac{d^2}{dq^2}+
V_l(q)\right)\varphi_{0}(q)=\varepsilon_{0}\varphi_{0}(q) \, ,
\end{equation}
where $\varepsilon_{0}$ is the lowest eigenvalues of the adiabatic
Hamiltonian defined by the first equality.

Once this procedure is carried out for every value of the rotation
angle $\phi$, the BP of the ground state is obtained as
\begin{eqnarray}\label{bp}
    \gamma&=&i\int_cd\phi\left\langle\psi_{tot}\left|\frac{d}{d\phi}\right|\psi_{tot}\right\rangle\nonumber\\
    &=&\int_{-\infty}^{+\infty}dq\varphi_0^2(q)\int_0^{2\pi}d\phi A(q,\phi)
\end{eqnarray}
where the $q$-parametrized connection is
\begin{eqnarray}\label{bp2}
    A(q,\phi)&=&i\left\langle\chi_l(q,\phi)\left|\frac{d}{d\phi}\right|\chi_l(q,\phi)\right\rangle=
    -{N}\frac{d\zeta}{d\phi}\cos^2{\frac{\beta}{2}}\nonumber\\
     &=&-\frac{ND}{2E(q)}\frac{\frac{L q}{\sqrt{N}}\cos{\phi}}{E(q)-\frac{L q}{\sqrt{N}}\cos{\phi}}
\end{eqnarray}
Substituting this expression into Eq.(\ref{bp}), straightforward
calculations lead to
\begin{equation}\label{bp3}
    \gamma=N\pi\left(1+\frac{\langle S_x\rangle}{N}\right)
\end{equation}
where
\begin{equation}\label{sxm}
\frac{\langle S_x\rangle}{N}=-\int_{-\infty}^\infty
\varphi_{0}^2(q) \frac{D}{E(q)} dq \, ,
\end{equation}

\begin{figure}
 \includegraphics{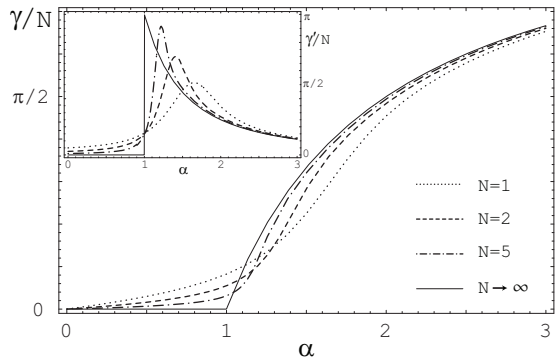}\\
 \caption{\label{berry} Numerical results for the scaled BP as
a function of the parameter $\alpha$, for $D=10$ and for different
values of $N$, in comparison with the result for
$N\rightarrow\infty$. The BP increases with the coupling, and, in
the thermodynamic limit, it displays a cusp at the critical value
$\alpha=1$. The inset shows the derivative of the BP with respect
to $\alpha$.}
\end{figure}

In the thermodynamic limit, one can show \cite{finite} that
\begin{equation}
 \frac{\langle S_x\rangle}{N}=\left\{%
\begin{array}{ll}
    -1 & \hbox{$(\alpha\leq 1)$} \\
    \hbox{$-\frac{1}{\alpha}$} & \hbox{$(\alpha> 1),$} \\
\end{array}%
\right.
\end{equation}
and thus, for $N\rightarrow \infty$, the BP is given by
\begin{equation}
 \frac{\gamma}{N} \Bigr |_{N\rightarrow \infty} =\left\{%
\begin{array}{ll}
    0 & \hbox{$(\alpha\leq 1)$} \\
    \hbox{$\pi(1-\frac{1}{\alpha})$} & \hbox{$(\alpha> 1)$} \\
\end{array}%
\right.
\end{equation}
Numerical results for the scaled BP are plotted in
Fig.(\ref{berry}) as a function of the parameter $\alpha$, for
$D=10$ and for different values of $N$, in comparison with the
result for the thermodynamic limit.
\begin{figure}
\includegraphics[width=0.3\textwidth]{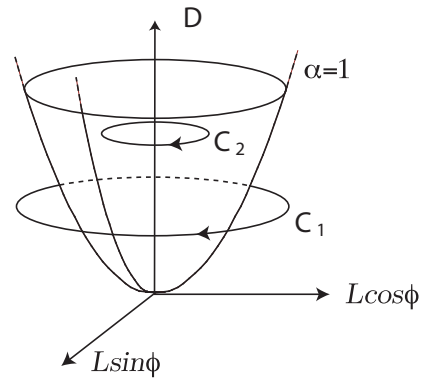}\\
\caption{\label{parab} A qualitative illustration of the paths
followed by the parameters of the Hamiltonian due to the
application of $U(\phi)$. The paraboloid corresponds to the value
$\alpha=L^2/2D=1$, for which the Hamiltonian shows a critical
behavior. If the parameters follow a path, e.g. $C_1$, encircling
the paraboloid, then the system acquires a non-trivial BP, which
tends to $\pi$ for $\alpha\gg 1$. As seen from figure~\ref{berry},
path $C_2$ gives rise to a zero Berry phase (in the
thermodynamical limit).}
\end{figure}
One can see that the BP increases with increasing the coupling
strength and, in the thermodynamic limit, its derivative becomes
discontinuous at the critical value $\alpha=\alpha_c=1$. This
results are in accordance with the expected behavior of the BP
across the critical point. Notice that, in the thermodynamic
limit, a non-trivial Berry phase is only obtained when a region of
criticality is encircled, as for the path $C_1$ in
Fig.~\ref{parab}. Indeed, in the enlarged parameter space
generated by the application of the unitary operator $U(\phi)$ of
Eq. (\ref{ut}), the critical point corresponds to the paraboloid
$\alpha = \frac{L^2}{2 D} =1$. As the radius of the path is
determined by $\alpha$, one can see that, in the limit
$N\rightarrow \infty$, the BP is zero in the normal phase ($\alpha
\leq 1$) and is non-zero in the super-radiant phase, i.e. if the
path encloses the critical region.

We now investigate the scaling behavior of BP at the critical
point by a finite size scaling approach. In order to obtain an
analytic estimation of BP as a function of $N$, we expand the
adiabatic potential in Eq.(\ref{se}) and obtain an anharmonic
oscillator potential
\begin{equation}\label{adpot}
    U_l(q)=\frac{2}{\omega}V_l(q)\simeq -ND+(1-\alpha)q^2+\frac{\alpha^2}{2ND}q^4
\end{equation}
The eigenvalue problem defined by this potential can be solved
with the help of Symanzik scaling \cite{finite,simon}. This is
done by rewriting Eq.(\ref{se}) into the equivalent form
\begin{equation}\label{hamscaling}\left[-\frac{d^2}{dx^2}+\zeta
x^2+x^4\right]\varphi_{0}(x;\zeta)
=e_{0}\left(\zeta\right)\varphi_{0}(x;\zeta)
\end{equation}
where the scaled position is $x= q \left ( \frac{\alpha^2}{2 ND}
\right)^{1/6}$, while
$\zeta=\left(\frac{2ND}{\alpha^2}\right)^{2/3}(\alpha_c-\alpha)$.
Finally, the energies in Eq.(\ref{se}) and (\ref{hamscaling}) obey
the scaling relation
\begin{equation}\label{sr}
\frac{2}{\omega}
\varepsilon_{0}(\alpha,ND)=-ND+\left(\frac{\alpha^2}{2ND}\right)^{1/3}e_{0}
\left(\zeta\right).
\end{equation} Since $\zeta \rightarrow 0$ at
the critical point, we can consider the $x^2$ term to be a
perturbation and employ the Rayleigh-Schr\"{o}dinger perturbation
theory. This yields the expansion $ e_{0}(\zeta)=\sum_{n=0}^\infty
c_n\zeta^n $, where  the coefficients $c_n$ can be obtained after
solving the equation for a purely quartic oscillator. It is easy
to get $c_0= e_{0}(0)\simeq1.06036$ and $c_1=\int_{-\infty}^\infty
q^2\phi_{0}^2(q;0)dq=e_0^\prime(0)\simeq 0.36203$.
\begin{figure}
 \includegraphics{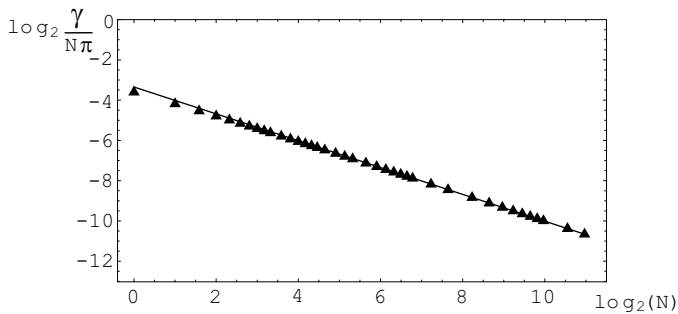}\\
 \caption{\label{gafn} Scaling of the BP
 as a function of $N$ at the critical point $\alpha=1$, for
 $D=10$. For easy of comparison, the continuous plot shows
the analytic expression of Eq. (\ref{BPscal}).}
\end{figure}

It can be shown that the coefficients of this expansion completely
determine the average value of every physical observable at the
critical point \cite{finite}. In particular, a similar expansion
applied to $\langle S_x \rangle$, allows us to write
\begin{equation}\label{sxsca}
\frac{\langle
S_x\rangle}{N}\simeq-1+\frac{2c_1}{(2ND)^{2/3}}-\frac{2c_0}{(2ND)^{4/3}}.
\end{equation}
Thus, we obtain the leading orders in the finite size scaling of
the Berry phase as
\begin{equation}\label{BPscal}
    \frac{\gamma}{N}\approx\pi\left[\frac{2 c_1}{(2ND)^{2/3}}-\frac{2
    c_0}{(2ND)^{4/3}}\right].
\end{equation}
This expression shows how the scaled geometric phase goes to zero
as $N$ increases and how the singular thermodynamic behavior is
approached at $\alpha=\alpha_c=1$. This analytical result, and in
particular the leading critical behavior $\gamma/N \sim  N^{-2/3}$
is confirmed in Fig. (\ref{gafn}) by comparison with the BP
obtained numerically from Eqs. (\ref{bp3})-(\ref{sxm}). In fact,
including also the second order correction scaling as $N^{-4/3}$,
we are able to reproduce the numerical result even for small
values of $N$.

Besides the scaling relation at the critical point $\alpha=1$, we
can also obtain the leading $1/N$ correction to the thermodynamic
limit of $\frac{\gamma}{N}$ for small and large values of
$\alpha$. Using the fact that the oscillator localizes around
$q=0$ for $\alpha \ll1$, while its wave function is split in two
components peaked around $\pm q_m$ for $\alpha \gg 1$, we get
\begin{equation}
\frac{\gamma}{N} -  \frac{\gamma}{N} \Bigr |_{N\rightarrow \infty} \approx \left\{%
\begin{array}{ll}
    \hbox{$\frac{\pi \alpha}{2 ND}$} & \hbox{$(\alpha\ll 1)$} \\
    \hbox{$- \frac{\pi}{ND \alpha^2}$} & \hbox{$(\alpha \gg 1)$} \\
\end{array}%
\right.
\end{equation}

To conclude, we have shown that the behavior of the geometric
phase in correspondence of the critical region for the Dicke
Model, confirms an expected connection between BP and QPT. Indeed,
BP and QPT share the common feature of both appearing in presence
of a singularity in the energy density of the system. This
heuristic argument motivates the need to explore the use of BP as
a tool to signal and investigate critical features of certain
models. However, strictly speaking, singularities in the energy
density of many body systems only appear in the thermodynamic
limit. It is therefore not obvious that in a \emph{finite scale}
regime, such a connection between BP and QPT can still be drawn.
Studying the BP in this regime has clearly theoretical interest
and obvious experimental motivations. In the case of the Dicke
model, we have found that the geometric phase shows, at finite
sizes, a precursor of the topological character which appears in
the thermodynamic limit. Moreover, studying the scaling of the BP
as a function of the system size, we have identified its critical
exponent.

%
%%%%%%%%%%%%%%%%%%%%%%%%%%%%%%%%%%%%%%%%%%%%%%%%%%%%%%%%%%%%%%%%%

\end{document}